\documentclass[prb,twocolumn,showpacs,preprintnumbers,amsmath,amssymb]{revtex4}
\usepackage{amssymb} \usepackage{graphicx} \usepackage{amsmath}
\begin{document}
\title{One-step replica symmetry breaking solution for a highly asymmetric two-sublattice fermionic Ising spin glass model in a transverse field} 
\author{F. M. Zimmer} \author{S.G. Magalhaes}\email{ggarcia@ccne.ufsm.br} \affiliation{Laborat\'orio de Mec\^anica Estat\'\i stica e Teoria da Mat\'eria Condensada, Universidade Federal de Santa Maria, 97105-900 Santa Maria, RS, Brazil}%

\begin{abstract}
The  one-step replica symmetry breaking (RSB) is used to study a two-sublattice fermionic 
infinite-range Ising spin glass (SG) model in a transverse field $\Gamma$.
The problem is formulated in a Grassmann path integral formalism within the static approximation.
In this model, a parallel magnetic field $H$ breaks the symmetry of the sublattices. It destroys the antiferromagnetic (AF) order, but it can favor the nonergodic mixed phase (SG+AF)
characterizing an asymmetric RSB region. 
In this region, intra-sublattice disordered interactions $V$ increase the difference between the RSB solutions of each sublattice.
The freezing temperature shows a higher increase with $H$ when $V$ enhances.
A discontinue phase transition from the replica symmetry (RS) solution to the RSB solution can appear with the presence of an intra-sublattice 
ferromagnetic average coupling.
The $\Gamma$ field introduces a quantum spin flip mechanism that suppresses the magnetic orders leading them to quantum critical points.
Results suggest that the quantum effects are not able to restore the RS solution.
However, in the asymmetric RSB region,
$\Gamma$ can produce a stable RS solution at any finite temperature for a particular sublattice while the other sublattice still presents RSB solution 
for the special case in which 
only the intra-sublattice spins couple with disordered interactions.

\end{abstract}
\maketitle

\section{Introduction}

The quantum spin glass (SG) theory\cite{braymoore} has been subject of extensive studies mainly by 
its conceptual difficulties to describe the ordered phase.
This problem becomes even more interesting when the SG and a long range antiferromagnetic (AF) interaction are both considered. 
The glassy AF order can be source of complex physics processes. For instance, 
quantum fluctuations and the interplay among disorder, frustrated magnetism and other interactions have 
been an essential feature of some heavy fermion systems\cite{experimental1} and high temperature superconductors.\cite{experimental2}
It can lead to a non-Fermi liquid behavior and to a quantum critical point (QCP). 
In this case, it is important to establish a theoretical framework to treat the SG/AF competition with other physical processes such as Kondo effect\cite{teorico1} or superconductivity\cite{teorico2} in the same footing at low temperature in which quantum effects become significant. We do not intend discussing this specific problem, but we concentrate on   the quantum magnetic properties of a fermionic formulation for a two-sublattice transverse field Ising SG model.

The infinite range Sherrington-Kirkpatrick SK model\cite{sk} in a transverse tunneling field $\Gamma$ has been extensively studied as a quantum SG model that presents a QCP 
and an experimental realization in the system LiHo$_x$Y$_{1-x}$F$_4$.\cite{experimental}  
However, even this simplest theoretical quantum SG model has recently been theme of controversial results concerning the local stability of replica symmetry (RS) solution (see Refs. \onlinecite{thirumalai,trotter,eduardo,zimmerprb}).
For instance, the classical SK model is well described by Parisi's replica symmetry breaking (RSB) solution.\cite{parisi} It considers the SG phase as a nonergodic situation in which the free energy landscape is characterized with many valleys separated by large barriers. 
This raises the question of whether or not the quantum tunneling through these barriers is able to restore the RS when quantum effects are included ($\Gamma>0$).

Concerning the SG/AF competition, a two-sublattice Ising spin model with SK-type interactions between classical spin variables of distinct sublattices has been introduced by Korenblit and Shender (KS) to discuss the  SG/AF problem at mean field level.\cite{ks}
A generalization of the KS model (GKS model) allows inter- and intra-sublattice disordered interactions.\cite{classico,takayama}
These two models (KS and GKS) have shown an interesting feature related to a parallel magnetic field $H$. It breaks the symmetry between the sublattices that can favor frustration increasing the RSB temperature range.\cite{ks,classico,takayama} 
The RSB with this asymmetry caused by the $H$ field is named here as asymmetric RSB region,
 in which the SG and the usual AF order coexist (mixed phase). 
However, a quantum version for the GKS model has received less consideration.
In this case, the quantum effects on the asymmetric RSB region remain an open question. 
For instance, the RSB occurs at the same temperature  in both sublattices for the classical GKS model.\cite{classico}
However, the degree of RSB can be distinct in each sublattice within the asymmetric RSB region. Regarding that difference, we can ask how the RSB behavior is when quantum fluctuations are present in the GKS model.
 
On the other hand, there is a particular situation for the classical GKS model (PGKS) in which the inter-sublattice interactions are only antiferromagnetic without disorder and the intra-sublattice spins only couple with disordered interactions.\cite{classicoab} 
In this case, the RSB can first occur for a specific sublattice (let us say $b$), while the other sublattice ($a$) still presents RS solution.
The sublattice $a$ can also show a RSB solution, but at a lower temperature. 
Now, the quantum tunneling is more pronounced at lower temperature. 
 In this context, the RSB solution of the sublattice $a$ can be more affected by the quantum fluctuations than the sublattice $b$. In this case, could  the sublattice $a$ present a stable RS  solution for any finite temperature at the same time that the sublattice $b$ still shows a RSB solution when $\Gamma>0$?
That is an important point for the SG/AF problem  because the solution of a particular sublattice also depends on the order parameters of the other sublattice.\cite{ks,classico} 
 Moreover, the frustration in the asymmetric RSB region could be changed if only the spins of the sublattice $b$ present RSB solution.

The aim of this paper is to investigate the effects of a transverse magnetic field $\Gamma$ on a fermionic  formulation for the GKS model where inter- and intra-sublattice SK-type interactions are considered. 
In special, we investigate 
the effects produced by the $\Gamma$ field on the frustration in the asymmetric RSB region for the case in which only the intra-sublattice interactions are disordered (PGSK model).
The Parisi's scheme \cite{parisi} of one-step RSB (1S-RSB) is used to study the non-trivial ergodicity breaking region.  In this case, we check 
if the quantum fluctuations are strong enough to cause tunneling between the free energy barriers of the many degenerated thermodynamic states.

The $H$ field breaks the symmetry between the sublattices in the GKS model.\cite{classico} As a consequence, $H$ can favor frustration at the same time that it destroys the AF order, so the freezing  temperature $T_f$ can increase with $H$. 
On the other hand, the $\Gamma$ field introduces a quantum spin flipping mechanism, which can suppress the magnetic orders leading their critical temperatures to QCPs.\cite{alba2002}
Therefore, both magnetic fields produce effects that can compete on $T_f$ in this two-sublattice SG/AF problem.
For example, in a previous fermionic version for the KS model in a transverse field (without intra-sublattice interaction),\cite{zimmer}  results within the RS solution have shown that  the effects produced by $\Gamma$ are dominant at low $H$. However, when $H$ enhances, the asymmetry between the sublattices becomes important and $T_f$ increases.
This scenario can be even more complex when the intra-sublattice disordered interactions are  considered. They could affect the $T_f$ behavior in the asymmetric RSB region. Furthermore,
a first-order phase transition from the AF to the paramagnetic phase can raise when $H$ increases.\cite{classico}
Therefore, in this paper, we also study the interplay between disorder and magnetic fields ($H$, $\Gamma$) in the SG/AF competition for the fermionic formulation of the GKS model in a transverse field  within 1S-RSB solution.

In this work, the hierarchical tree-structure  
is adopted in each sublattice
to obtain a RSB solution.\cite{parisi}
Thus, the elements of the replica matrices for the sublattices $a$ and $b$ are parameterized following the Parisi scheme of 1S-RSB.
These replica matrices are independent of each other for the PGKS model in which 
the two sublattices present independent 1S-RSB solutions. 
However, the replica matrices for the sublattices $a$ and $b$ are coupled for the GKS model.
In this case, we adopt the approximation $m_a=m_b$ in order to decouple them, where $m_p$ ($p=a$ or $b$) is the  Parisi block-size parameter\cite{parisi} for the sublattice $p$. 
In this approximation, the situation $m_a\neq m_b$ never occurs. 
Nevertheless, it still allows sublattice dependence for the behavior of the other 1S-RSB order parameters. 
An alternative RSB method based on the Parisi's scheme  and a wave-like structure has been proposed in Ref. \onlinecite{oppermann} for the KS model without intra-sublattice disorder. 
It is the modulated RSB scheme that considers a possible asymmetry between $m_a$ and $m_b$ which is controlled by a variational parameter in the replica limit.
However, this RSB scheme increases the numerical efforts to obtain the 1S-RSB results and it can only provide a very small improvement in the free energy compared with the Parisi scheme. Therefore, we do not use the modulated RSB scheme in the present GSK model, in which the intra-sublattice SK-type interactions add some difficulties when this RSB scheme is implemented.

The partition function of the present work is computed with the path integral formalism in which the spin operators are represented by bilinear combinations of Grassmann variables as in Ref. \onlinecite{alba2002}.
The 1S-RSB free energy is calculated within the static approximation (SA) that is used to treat the spin-spin correlation functions.\cite{braymoore} 
The SA deserves some comments. It is reasonable to obtain phase boundaries even at low temperature.\cite{sachdev,miller} 
However, it can yield inaccurate quantitative results 
for the dynamical behavior of the correlation functions
at very low temperature when $\Gamma>0$. 
Nevertheless, there are several recent papers\cite{alba2002,zimmer,eduardo,zimmerprb} adopting this particular approximation because of a lack of a suitable 
analytical method to go beyond the SA in the whole SG phase in the femionic formalism, in which the SG phase 
has also a shortcoming related to the RSB solution. Furthermore, when temperature decreases for small and intermediary values of $\Gamma$ in the one-lattice SG problem,\cite{zimmerprb} the results produced by SA with 1S-RSB are in good qualitative agreement with that obtained by Trotter-Suzuki formalism in which numerical methods are used to treat the dynamic of the spin-spin correlations.\cite{trotter}
Moreover, a Landau-Ginzburg formalism has been proposed in Ref. \onlinecite{rotores} to analyze the SG of quantum rotors near the QCP. In this formalism, the RSB occurs in the SG phase at finite temperature $T$ and it is suppressed when $T\rightarrow 0$ for the region close to the QCP. The results with SA agree qualitatively with this RSB behavior when $\Gamma$ increases towards the QCP, in which the RSB effects become less pronounced.\cite{zimmerprb}

This paper is structured in the following way.
In the next section, the model is introduced, and the 1S-RSB free energy is obtained. In section \ref{results}, numerical results for the  1S-RSB order parameters, free energy and phase diagrams showing the SG/AF competition in temperature versus parallel magnetic field are reported for several values of disorder and transverse magnetic field.
The last section is left for summary and conclusions. 

\section{General Formulation}\label{model} 
Let us consider a set of localized Ising spins at sites of two identical sublattices $a$ and $b$. The spins of same and distinct sublattices can interact with infinite-range random couplings. Transverse ($\Gamma$) and parallel ($H$) magnetic fields are coupled with spins of both sublattices. The model is represented by the Hamiltonian
\begin{equation}
\begin{split}
{\hat{ H}}= -\sum_{i_{a}j_{b}} J_{i_{a}j_{b}}\hat{S}_{i_{a}}^{z} \hat{S}_{j_{b}}^{z}
-\sum_{p=a,b}\sum_{i_{p}j_{p}} V_{i_{p}j_{p}}\hat{S}_{i_{p}}^{z} \hat{S}_{j_{p}}^{z}
\\
-2\sum_{i_a} \left(\Gamma\hat{S}_{i_{a}}^{x}+H \hat{S}_{i_{a}}^{z} \right)
-2\sum_{j_b} \left(\Gamma\hat{S}_{j_{b}}^{x}+H \hat{S}_{j_{b}}^{z} \right)
\label{ham}
\end{split}
\end{equation}
\noindent
where the sums run over the $N$ sites of each sublattice $p$ ($=a$ or $b$).
The inter- and intra-sublattice exchange interactions $J_{i_aj_b}$ and $V_{i_pj_p}$ are random variables that follow Gaussian distributions with mean values $-4J_0/N$ and  $2V_0/N$ and variances $J^2/32N$ and $V^2/16N$, respectively.
The spin operators in Eq. (\ref{ham}) are defined in terms of fermion operators \cite{alba2002}:
\begin{equation}
\hat{S}_{i_{p}}^{z}=\frac{1}{2}(\hat{n}_{i_p\uparrow}-\hat{n}_{i_p\downarrow})~
\mbox{and}~
\hat{S}_{i_{p}}^{x}=\frac{1}{2}(c_{i_p\uparrow}^{\dagger}c_{i_p\downarrow}+
c_{i_p\downarrow}^{\dagger}c_{i_p\uparrow})
\end{equation}
where $\hat{n}_{i_p\sigma}=c_{i_p\sigma}^{\dagger}c_{i_p\sigma}$ gives the number of fermions at site $i_p$ with spin projection $\sigma=\uparrow$ or $\downarrow$.
$c_{i_p\sigma}^{\dagger}$ and $c_{i_p\sigma}$ are the fermions creation and annihilation operators, respectively. In this representation,
the operator $S^{z}$ has two magnetic eigenvalues
and two nonmagnetic eigenvalues (empty and double occupied site). 
Therefore, there are two states that do not belong
to the usual spin space.
The present fermionic SG/AF problem admits these four states. However, the chemical potential $\mu$ is chosen to ensure the average occupation of one fermion per site.\cite{alba84} 

Lagrangian path integral formalism with spin operators represented as anticommuting Grassmann field ($\phi,~ \phi^*$) is used to treat this fermionic problem. \cite{alba2002, zimmer} Therefore, the partition function is written as: 
\begin{equation}
Z=\int D(\phi^* \phi)\mbox{e}^{A(\phi^*, \phi)}
\end{equation}
with the action
\begin{equation}
\begin{split}
A(\phi^*, \phi)=\int_{0}^{\beta}d\tau
\{\sum_{p=a,b}\sum_{j_p,\sigma}\phi_{j_p\sigma}^{*}(\tau)
 \frac{\partial}{\partial \tau}
 \phi_{j_p\sigma}(\tau)
 \\
  - H\left(\phi_{j_p\sigma}^{*}
 (\tau),\phi_{j_p\sigma}(\tau)\right) \}
\end{split}
\label{action},
\end{equation}
$\beta=1/T$ ($T$ is the temperature) and $\mu=0$, which corresponds to the half-filling situation. 

In Eq. (\ref{action}), the Fourier decomposition of the time-dependent quantities is employed. 
The average over the quenched disorders is performed by means of replica method ($\beta F
=-\frac{1}{2N}\displaystyle\lim_{n\rightarrow 0}(\langle Z^n \rangle_{J_{ij}} - 1)/n$)
with the standard Hubbard-Stratonovich transformations used to decouple different sites. Within the static ansatz, this procedure results in an effective single-site problem that involves saddle-point conditions for each 
sublattice 
magnetization $M_p$, replica diagonal spin-spin correlation $r_p$, replica off-diagonal elements $q_p^{\alpha,\gamma}$ which are related to the SG order parameters, and also fields $Q_3^{\alpha\gamma} (=(q_a^{\alpha\gamma}+q_b^{\alpha\gamma})/2)$ that couple the sublattices. 
Therefore, the replicated partition function is
\begin{equation}
\begin{split}
Z(n)=\beta\sum_{\alpha}\left[-J_0 M_{a}^{\alpha}M_{b}^{\alpha} +\frac{V_0}{2}\sum_{p}(M_{p}^{\alpha})^2\right]
\\
+\beta^2\sum_{\alpha,\gamma}\left[J^2 q_{a}^{\alpha\gamma}q_{b}^{\alpha\gamma} +\frac{V^2}{2}\sum_{p}(q_{p}^{\alpha\gamma})^2\right]
-\sum_{p}\ln \Theta_{p}
\label{saddle}
\end{split}
\end{equation}
where $\alpha$ and $\gamma$ are the replica indices running from 1 to $n$,
\begin{equation}
\Theta_{p}
=
\int D[\phi_p^{\alpha *}\phi_p^{\alpha}]\exp\left[H_{p}^{eff}\right]
\label{lambdaeff}
\end{equation}
\noindent
with
\begin{equation}
\begin{split}
H_{p}^{eff}= 
\sum_{\alpha} \left[ A_{M,p}^{\alpha} + 2\beta\left(V_0M_p^{\alpha}-J_0 M_{p'}^{\alpha} \right)S_{p}^{\alpha}\right] 
\\
+ 
4\beta^2 \sum_{\alpha\gamma}\left[V^2 q_{p}^{\alpha\gamma} + J^2 q_{p'}^{\alpha\gamma}\right]S_{p}^{\alpha}S_{p}^{\gamma}
\label{heff},
\end{split}
\end{equation}
\noindent 
and $p=a~(p'=b)$ or $p=b~ (p'=a)$. In Eq. (\ref{heff}), 
\begin{equation} 
A_{M,p}^{\alpha}=\sum_{\omega}\underline{\Phi}_{p}^{\alpha\dagger}(\omega)(i\omega+
\beta H\underline{\sigma}^{z}+\beta\Gamma\underline{\sigma}^{x})
\underline{\Phi}_{p}^{\alpha}(\omega),
\end{equation} 
\begin{equation} 
\\
S_{p}^{\alpha}=\frac{1}{2}\sum_{\omega}
\underline{\Phi}_p^{\alpha\dagger}(\omega)\underline{\sigma}^{z}\underline{\Phi}_p^\alpha(\omega),
\end{equation}
$\underline{\Phi}_p^{\alpha\dagger}(\omega)=\left(\phi_{p\uparrow}^{\alpha *}(\omega)~~~
\phi_{p\downarrow}^{\alpha *}(\omega)\right)$ is a Grassmann spinor, $\underline{\sigma}^x$ 
and $\underline{\sigma}^z$ are the Pauli matrices, and $\omega$ denotes fermionic Matsubara 
frequencies.

The Parisi's scheme of one-step replica symmetry breaking (1S-RSB) is used to parametrize  $q_{p}^{\alpha\gamma}$ as\cite{parisi}
\begin{equation}
q_{p}^{\alpha\gamma}=\left\{ 
\begin{aligned}
r_p &\mbox{ if } \alpha=\gamma
\\
q_{p1} &\mbox{ if } I(\alpha/m_{p})=I(\gamma/m_{p})
\\
q_{p0} &\mbox{ if } I(\alpha/m_{p})\neq(\gamma/m_{p})
\end{aligned}\right.
\label{parametrization}
\end{equation}
where $I(x)$ gives the smallest integer which is greater than or equal to $x$.
 
For a general case, when $J$ and $V$ are non-zero (GKS model), the Parisi block-size parameter $m_p$ is assumed to be equal to $m$ 
for both sublattices. 
This approximation is used in order to follow the standard Parisi scheme of RSB. 
However, the situation with $J=0$ and $V\neq 0$ (PGKS model) is analyzed by taken $m_a$ and  $m_b$ 
as independent
parameters ($m_a\neq m_b$) 
in the next section. 

The parametrization (\ref{parametrization}) with $m=m_a=m_b$  is used to sum over the replica index. 
It produces quadratic terms in Eq. (\ref{heff}) that are linearized introducing new auxiliary fields in Eq. (\ref{lambdaeff}).
The functional Grassmann integral is now performed, and the sum over the Matsubara 
frequencies can be evaluated like Ref. \onlinecite{alba2002}.  
This procedure results in the following expression to the 1S-RSB free energy: 
\begin{equation}
\begin{split}
2 F=- J_{0} M_a M_b +
\beta J^2 \left[m(q_{a1}q_{b1}-q_{a0}q_{b0})+r_{a}r_{b}
\right.\\ \left. 
-q_{a1}q_{b1}\right]
+\sum_{p=a,b}\left\{
\frac{V_0}{2}M_p^2
+\frac{\beta V^2}{2}\left[m(q_{p1}^2-q_{p0}^{2})
\right. \right.\\ 
\left.
\left.+r_{p}^2-q_{p1}^2\right]
-
\frac{1}{\beta m}\int_{-\infty}^{\infty}Dz_p \ln \int_{-\infty}^{\infty}Dv_p\Theta_{p}^m\right\}
-\frac{\ln 4}{\beta}
\label{freeenergy}
\end{split}
\end{equation}
with 
\begin{equation}
\Theta_{p}=1+
\int_{-\infty}^{\infty}D\xi_p
\cosh(\beta\sqrt{\Delta_p})
\label{difference}
\end{equation}
where $\Delta_p=h_p^{2}+\Gamma^2$, $h_p=\bar{h}_p+\sigma_{1p}v_p+\sigma_{2p}$,  $\bar{h}_p=H+V_0M_p-J_0M_{p^{'}}$ ($p\neq p^{'}$),
\begin{equation}
\sigma_{1p}=\sqrt{2[V^2(q_{p1}-q_{p0})+J^2(q_{p^{'}1}-q_{p^{'}0})]}, 
\label{variance}\end{equation}
\begin{equation}
\begin{split}
\sigma_{2p}=
\sqrt{2[V^2(r_p-q_{p1})+J^2(r_{p^{'}}-q_{p^{'}1})]}\xi_p
\\
+\sqrt{2(V^2q_{p0}+J^2q_{p^{'}0})}z_p,
\label{variance2}\end{split} \end{equation}
and $Dx=dx~\mbox{e}^{-x^2/2}/\sqrt{2\pi}$ ($x=z_p, v_p \mbox{ or } \xi_p$).
The parameters $q_{p0}$,
$q_{p1}$, $M_p$, $r_p$, and $m$ are given by the extreme condition 
of free energy (\ref{freeenergy}).

The 1S-RSB solution of free energy (\ref{freeenergy}) recovers the RS one ($q_{p1}=q_{p0}$ and $m=0$) at temperature $T$ above the freezing temperature $T_f$. 
In this case, the stability against transversal fluctuations of the RS solution follows from the Almeida-Thouless analysis,\cite{almeida} which is extended to the two-sublattice problem.\cite{ks}
The 
$T_f$ 
is given by the higher temperature that satisfies the expression:
\begin{equation}
(T_{f}^{2}-V^2d_a)(T_{f}^{2}-V^2d_b)-J^4d_ad_b=0
\label{at}
\end{equation}
where
\begin{equation}
\begin{split}
d_{p}=2\int_{-\infty}^{\infty}Dz_p\left\{
\left(
\frac{\int D\xi_p\frac{h_{p}}{\sqrt{\Delta_{p}}}\sinh(\beta\sqrt{\Delta_{p}})}{\Theta_{p}}
\right)^{2}
\right.\\
\left.
-\frac{\int D\xi_p
[\frac{h_{p}^{2}}{\Delta_{p}}\cosh(\beta\sqrt{\Delta_{p}})+\frac{\Gamma^{2}}{\beta\Delta_{p}^{3/2}}
\sinh(\beta\sqrt{\Delta_{p}})
]}{\Theta_{p}} \right\}^2 
\end{split}
\end{equation}
with $\sigma_{1p}=0$ in $h_p$.

\begin{figure}[t!]
 \includegraphics[angle=270,width =8.5cm]{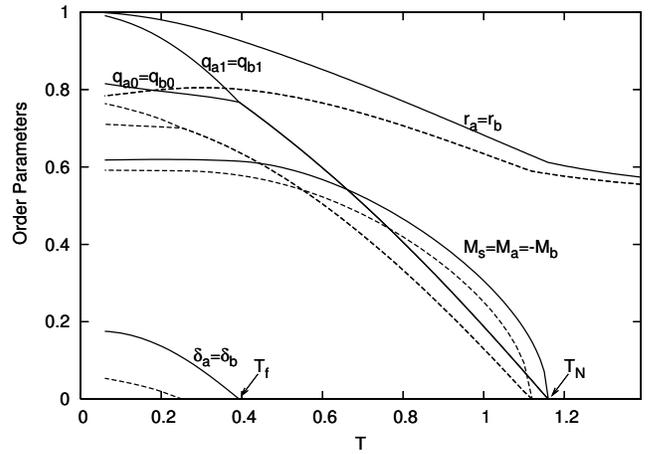} 
 \caption{Order parameters versus $T$ for $H=0$ and two values of $\Gamma$: $\Gamma=0$ (full lines) and $\Gamma=0.8$ (dashed lines). }
 \label{params4}
\end{figure}

\section{Results}\label{results}

Numerical calculations for the set of saddle-point equations show different types of solutions. These depend on the values of various parameters of the model. For instance, in the absence of parallel magnetic field $H$, the paramagnetic (PM: $q_{p1}=q_{p0}=0$, $M_p=0$),
the spin glass (SG: $q_{p1}\neq q_{p0}>0$, $M_p=0$) and the antiferromagnetic (AF: $q_{p1}=q_{p0}>0$, $M_a=-M_b>0$) solutions can be found. 
A mixed phase (SG+AF) can also be obtained when $M_a=-M_b>0$ with RSB ($q_{p1}\neq q_{p0}$) solution, which is described by the order parameter $\delta_p\equiv q_{p1}- q_{p0}>0$.   
The presence of $H$ always induces the parameters $q_{p1}$ and $q_{p0}$. In this context, the PM solution is characterized by $q_{p1}=q_{p0}$ and $M_a=M_b$. The AF solution is found when $M_{a}\neq M_b$ with $q_{p1}=q_{p0}$. Particularly, the AF order is given by the staggered magnetization $M_{s}\equiv (M_a-M_b)/2>0$ and $\delta_p=0$. 
The SG phase occurs when $M_s=0$ with RSB solution. The mixed phase is found when $M_s>0$ with $\delta_p>0$.

\begin{figure}[t!]
 \centering
 \includegraphics[angle=270,width =8.5cm]{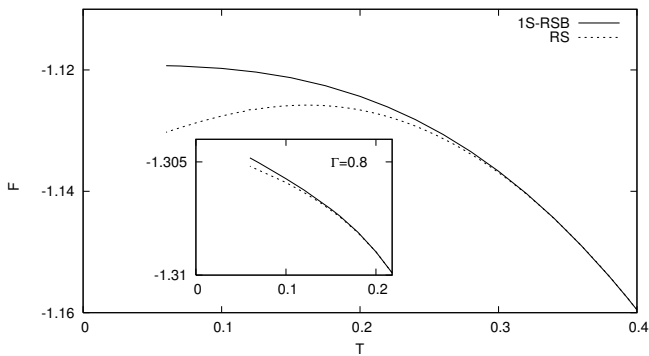} 
 \caption{Free energy versus $T$ for $H=0$
 and  $\Gamma=0$. The insert shows the result for $\Gamma=0.8$.} 
\label{params4free}
\end{figure}

We first present results for $H=0$. In this case, the order parameters as a function of $T$ are exhibited in Fig. (\ref{params4}) for 
a given degree of frustration $\eta\equiv \sqrt{V^2+J^2}/(V_0+J_0)$. In this work, the quantities $T$, $H$, $\Gamma$, $V_0$ and $J_0$ are in units of $\sqrt{V^2+J^2}$. For numerical purposes $\sqrt{V^2+J^2}=1$ and $\eta=1/1.9$.
The order parameter $M_s$ appears continuously at the Neel temperature $T_N$ when $T$ decreases. The RSB occurs at $T_f$, in which $\delta_p$ becomes greater than zero. At this point, the staggered magnetization is non zero,  and this (with $\delta_p>0$)  characterizes the mixed phase. 
The magnetization of each sublattice becomes weakly dependent on $T$ below $T_f$. 
Furthermore, when $H=0$, the two sublattices exhibit symmetric results for the order parameters, 
which are identical for any set of $V_0$, $J_0$, $V$ and $J$ that satisfies the same value of $\eta$.
The transverse field $\Gamma$ decreases the values of the order parameters as well as the critical temperatures (see dashed lines of Fig. (\ref{params4})). Its effects are more pronounced at low temperature. However, it is not able to restore the RS solution for $H=0$ that is in agreement with Refs. \onlinecite{trotter} and \onlinecite{zimmerprb}.  
The free energy of 1S-RSB solution can be compared with the one of the RS solution in Fig. (\ref{params4free}). 
The difference between these two approaches increases when $T$ decreases from $T_f$, and the entropy ($S=-\partial F/\partial T$) of the 1S-RSB solution is positive for the range of temperature analyzed. 

The results for $H>0$ can depend on the set of parameters $V_0$, $J_0$, $V$ and $J$.
However, it is not our intention to do an exhaustive study of all possible configuration of interactions for the model (\ref{ham}).
Nevertheless, we have chosen a specific degree of frustration in which the role of each parameter is analyzed.
In this case, the ratio $T_N(H=0)/T_f(H=0)$ remains the same for all results.
For example, we analyze the behavior of the RSB solution and the phase diagrams when the intra-sublattice ferromagnetic average coupling $V_0$ is changed, or when the intra-sublattice disorder ($V$) increases.
Moreover, the main focus of the work is to study the effects of quantum fluctuations on the asymmetric RSB region.
For this purpose, we consider two types of coupling: the first one (GKS model) admits SK-type interactions between spins of same and distinct sublattices ($J,~V\neq 0$). The second one (PGKS model) considers only an antiferromagnetic coupling ($J=0,~J_0>0$) between the sublattices 
with intra-sublattice SK-type interactions ($V\neq0$).

\subsection{Asymmetric RSB for $V\neq 0$ and $J\neq 0$ (GKS)}\label{resultaA}

Results of the 1S-RSB order parameters are discussed for the first type of coupling in the asymmetric RSB region.
To do so, we study
three distinct configurations of inter- and intra-sublattice interactions that satisfy the same $\eta$: (i) $J=V$ and $V_0=J_0=0.95$, (ii) $J=V$, $V_0=1.56$ and $J_0=0.34$, and (iii) $J=0.24$, $V=0.97$ and $V_0=J_0=0.95$.
In these three cases, we can analyze the main effect cased by  increasing $V_0/J_0$ (consider cases (i) and (ii)) and by increasing $V/J$ (consider cases (i) and (iii)).

\begin{figure}[t!]
 \includegraphics[angle=270,width =7.9cm]{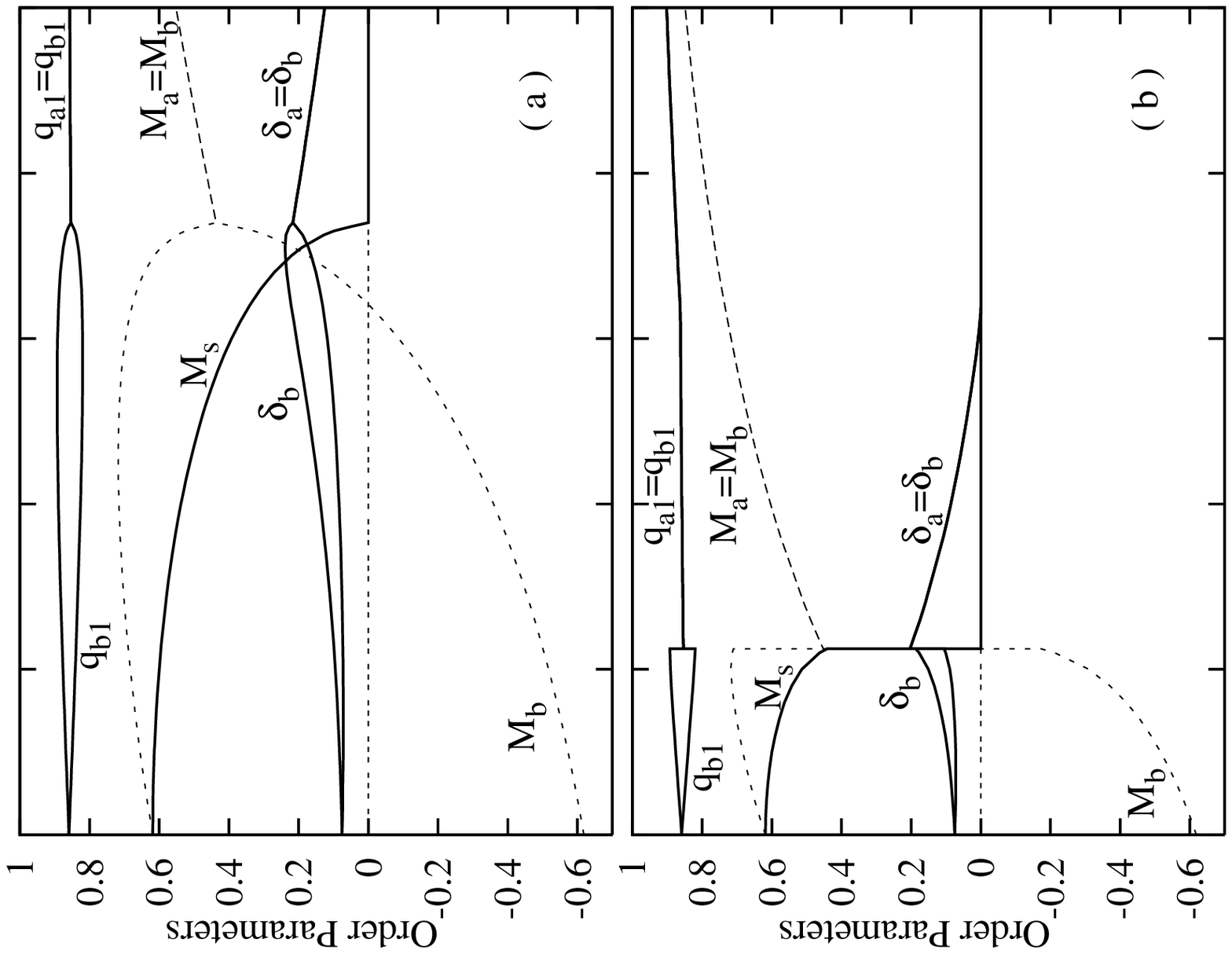} 
 \includegraphics[angle=270,width =7.9cm]{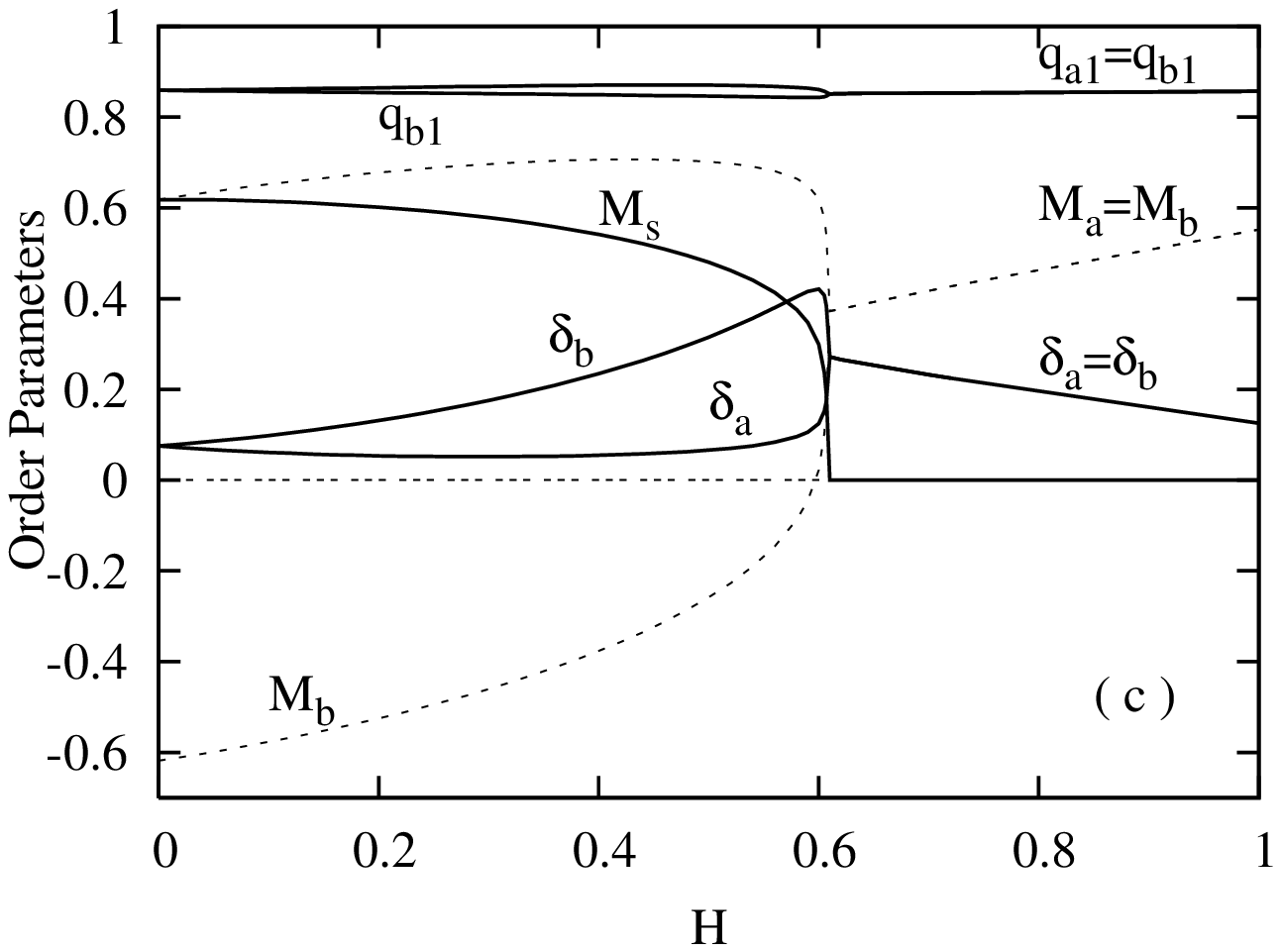} 
 \caption{Order parameters versus $H$ for $T=0.3$ and $\Gamma=0$. The results are for the following parameters: (a) $J=V$ and $V_0=J_0=0.95$, (b) $J=V$, $V_0=1.56$ and $J_0=0.34$, and (c) $J=0.24$, $V=0.97$ and $V_0=J_0=0.95$.}
\label{params4h}
\end{figure}

\begin{figure}[t!]
 \includegraphics[angle=270,width =8.5cm]{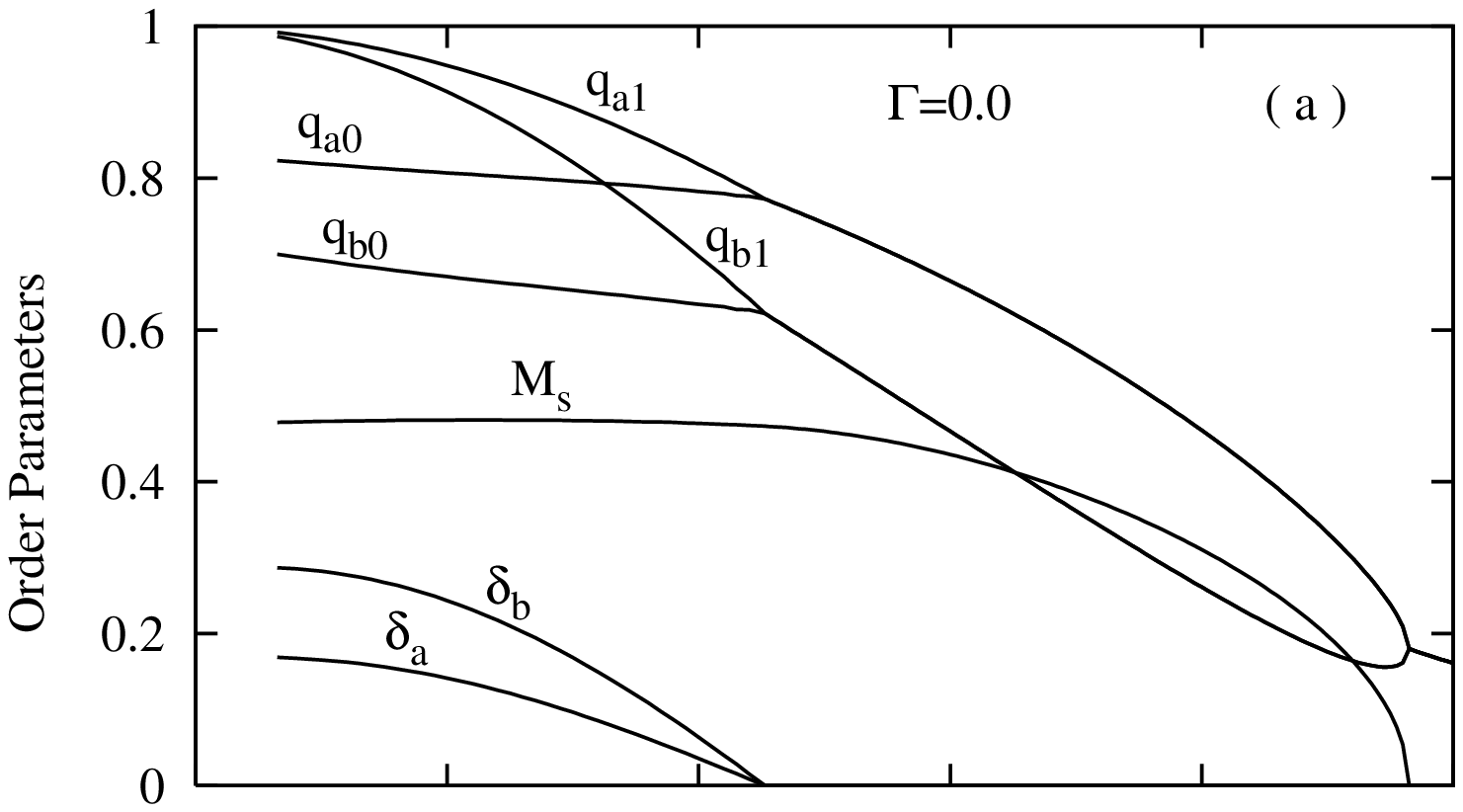} 
 \includegraphics[angle=270,width =8.5cm]{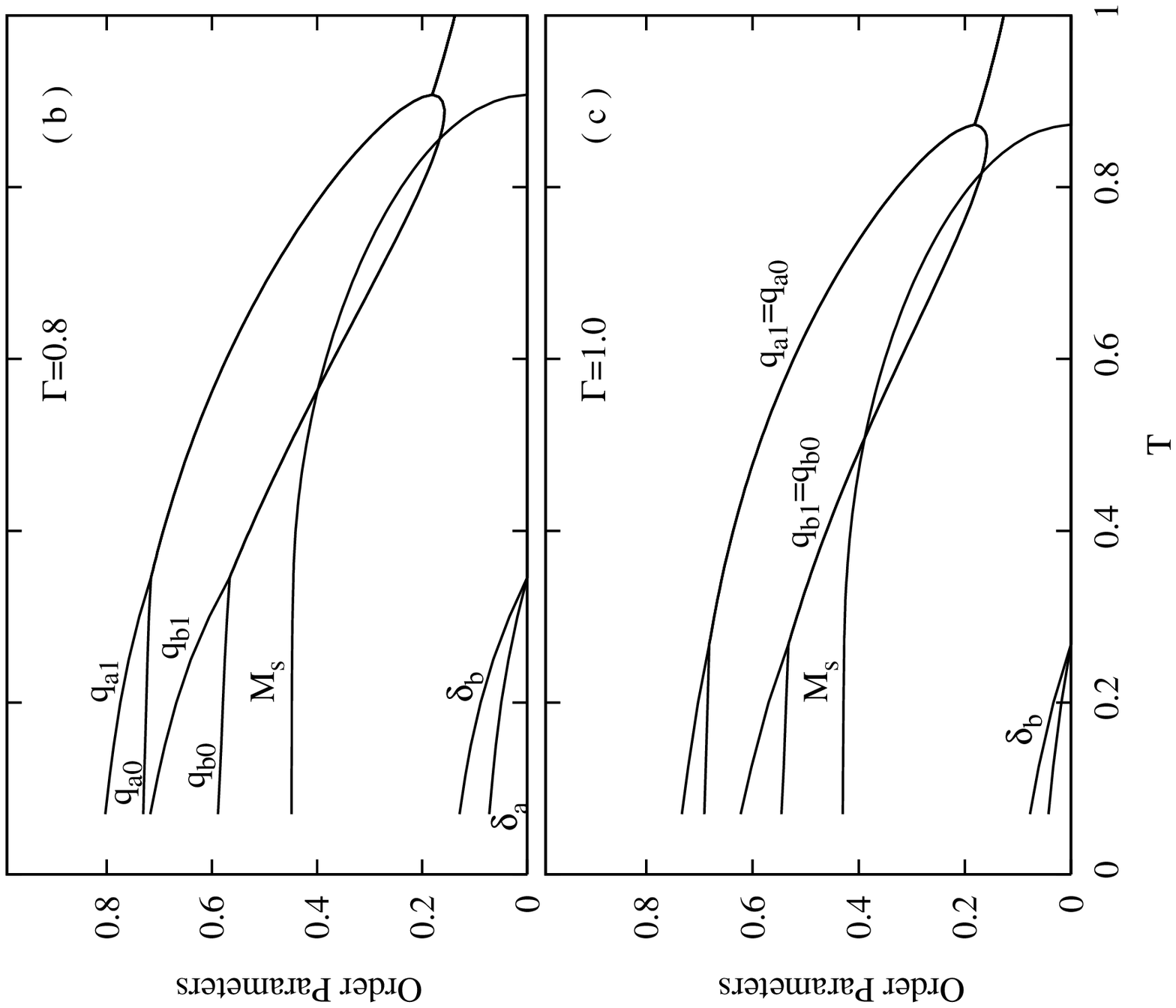} 
 \caption{Order parameters as a function of $T$ for $H=0.5$ and three values of $\Gamma$ (=0.0, 0.8 and 1.0). The interactions are given by case (i): $J=V$ and $V_0=J_0=0.95$.}
\label{paramga0810}
\end{figure}

The $H$-dependence of the order parameters $q_{p1}$, $\delta_{p1}$, $M_{p}$ and $M_s$ is reported in Fig. (\ref{params4h}) for $T=0.3$ and $\Gamma=0$. The sequence of panels (a), (b) and (c) in Fig. (\ref{params4h}) represents the cases (i), (ii) and (iii), respectively.
The $H$ field breaks the symmetry between the sublattices in the region where $M_s>0$.
However, the results for both sublattices are symmetric at high values of $H$ ($M_s=0$). 
An important point is the $H$-dependence of $\delta_p$.
For instance, $\delta_b$ can increase with $H$ in the mixed phase, which characterizes an enhancing of the RSB. 
This $\delta_b$ odd behavior can be associated qualitatively with the $H$-dependence of the average internal field $\bar{h}_b$ ($=H+V_0 M_b-J_0 M_{a}$) that acts on spins of sublattice $b$. 
For example, there is a region at small $H$ where the absolute value of magnetization $|M_b|$ decreases with $H$, while $M_a$ has a weaker dependence on $H$ than
$|M_b|$ (see Fig. (\ref{params4h})). As a consequence, $|\bar{h}_b|$ decreases with $H$ and $\delta_b$ enhances. 
Now, $\bar{h}_p$ is also related to $V_0/J_0$. When $V_0$ increases as in case (ii) (see Fig. (\ref{params4h}-b)), the region with solution $M_s>0$ decreases. The order parameters can show a discontinuity when $H$ crosses from the $M_s>0$ to the $M_s=0$ solution, where 
$\delta_p$ jumps to a higher value.
On the other hand, the relations in $\bar{h}_p$ are not directly affected by
the intra-sublattice disorder $V$.
However, $\delta_b$ is higher within the mixed phase when $V$ increases as it is shown by  comparing cases (i) and (iii) in Figs. (\ref{params4h}-a) and (\ref{params4h}-c).
For instance, the random internal field $h_p$, which acts on the sublattice $p$, also depends on the variance $\sigma_{1p}$ (see Eq. (\ref{variance})), which enhances (decreases) for the sublattice $b$ ($a$) in the asymmetric RSB region when $V/J$ increases.
Consequently, $\delta_b$ ($\delta_a$) can increase (decrease) with $V/J$ (Fig. (\ref{params4h}-c)). 
Therefore, the behavior of the 1S-RSB solution 
is also affected by the relation $V/J$.
In this case, the difference between the degree of RSB of the two sublattices is increased
with $V/J$, but the RSB always occurs in both sublattices simultaneous for $J\neq0$. 
In the next section, this result can be compared with the limit case $J=0$.

\begin{figure}[t!]
 \includegraphics[angle=270,width =8.3cm]{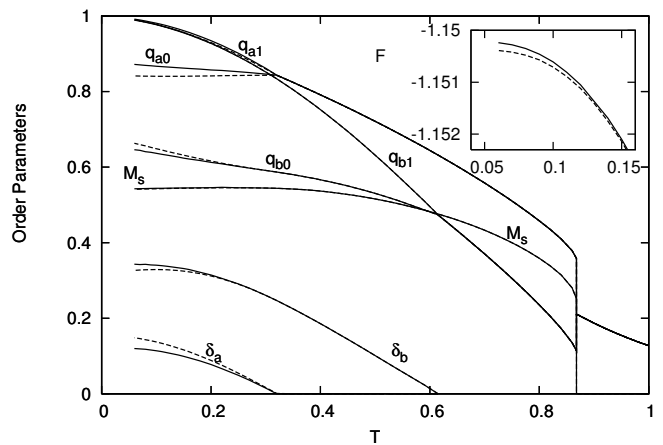} 
 \caption{Order parameters versus $T$ for $H=0.4$, $\Gamma=0.0$ and $J_0=V_0=0.95$, $J=0$ and $V=1.0$. The dashed lines represent results for the 1S-RSB solution when $m=m_a=m_b$.
 The insert shows the free energy versus $T$.}
 \label{paramj0ga0}
\end{figure}

Figure (\ref{paramga0810}) shows effects of quantum tunneling on the asymmetric RSB region. 
It exhibits the order parameters versus $T$ for the case (i) with $H=0.4$, $\Gamma=0.0$, $0.8$ and $1.0$.
The $\Gamma$ field decreases the values of order parameters $\delta_a$ and $\delta_b$, but 
the RSB still occurs at the same critical temperature for both sublattices. It means  that the quantum fluctuations act with the same intensity in the RSB solution of sublattices  $a$ and $b$, which still remain asymmetric.
Furthermore, 
the results suggest that the quantum tunneling is not able to restore the RS solution of neither of the sublattices, even in the asymmetric RSB region. 
The quantum effects on the order parameters for cases (ii) and (iii) are qualitatively the same as those obtained in case (i).

\subsection{Asymmetric RSB for $V\neq0$ and $J=0$ (PGKS)} \label{resultaB}

At this point, we restrict our study to the second type of coupling, in which the inter-sublattice exchange interactions are only antiferromagnetic without disorder and the intra-sublattice interactions are still disordered ($J_0\neq 0$, $J=0$ and $V\neq 0$).
In this case, the structures of the replica matrices $\{Q_a\}$ and $\{Q_b\}$ are independent of each other. 
The Parisi's scheme of RSB is applied in the two sublattices without any approximation on $m_p$. Therefore, the parameterization (\ref{parametrization}) is used to obtain the 1S-RSB free energy as in Eq. (\ref{freeenergy}). However, 
the parameters $m_a$ and $m_b$ are now both taken as variational parameters
of the free energy
\begin{equation}
\begin{split}
2 F=- J_{0} M_a M_b +\sum_{p=a,b}\left\{ \frac{V_0}{2}M_p^2
+\frac{\beta V^2}{2}
[m_p(q_{p1}^{2}-q_{p0}^{2})\right.
\\
+r_{p}^{2}
-q_{p1}^{2}]\left.-
\frac{1}{\beta m_p}\int_{-\infty}^{\infty}Dz_p \ln \int_{-\infty}^{\infty}Dv_p\Theta_{p}^{m_p}\right\}
-\frac{\ln 4}{\beta}
\label{freeenergy0}
\end{split}
\end{equation}
\noindent
with $\Theta_{p}$ defined in Eq. (\ref{difference}).
In this approximation $m_a\neq m_b$ in the asymmetric RSB region. 

\begin{figure}[t!]
\includegraphics[angle=270,width =8.3cm]{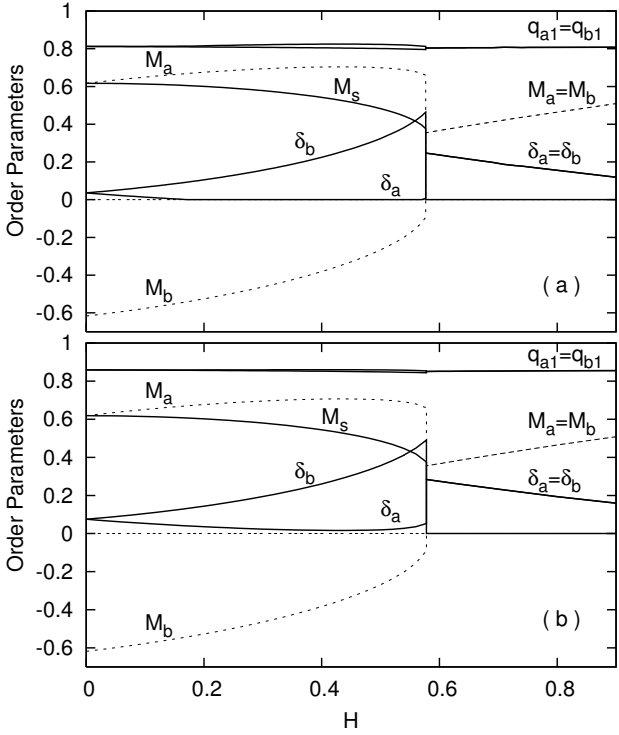} 
 \caption{Order parameters as a function of $H$ for $\Gamma=0.0$ and two values of $T$: (a) $T=0.35$ and (b) $T=0.3$. The disorder is the same as Fig. (\ref{paramj0ga0}).}
 \label{paramj0ga0t04}
\end{figure}

The RSB appears in both sublattices simultaneously with the same results as those obtained in Eq. (\ref{freeenergy}) in the symmetric region ($M_a=-M_b$ or $M_s=0$).
However, in the asymmetric region ($H>0$ and $M_s>0$),
the RSB occurs first at the sublattice $b$, while the sublattice $a$ can still present RS solution  as it is shown in Fig. (\ref{paramj0ga0}) for the particular interaction: $J_0=V_0=0.95$, $J=0$ and $V=1.0$ (case (iv)). 
Nevertheless, both sublattices present RSB solution at lower temperature when $\Gamma=0$.
In Fig. (\ref{paramj0ga0}), 
the mixed phase is characterized when 
 the replica symmetry is broken in the sublattice $b$.
However, the spin freezing effects are stronger when the two sublattices present RSB solution. For instance,
the staggered magnetization becomes less dependent on $T$ for the region where the sublattices $a$ and $b$ have RSB solution (see Fig. (\ref{paramj0ga0})). 
Therefore, we can identify two distinct asymmetric RSB regions: one with low number of  frustrated  spins where only the set of spins in the sublattice $b$ exhibits RSB solution; another with the RSB solution present in the two sublattices. It occurs because 
the internal fields that act on the sublattices $a$ and $b$ are different. 
In addition, the sublattices can exhibit frustration independently ($J=0$).

\begin{figure}[t!]
 \includegraphics[angle=270,width =8.3cm]{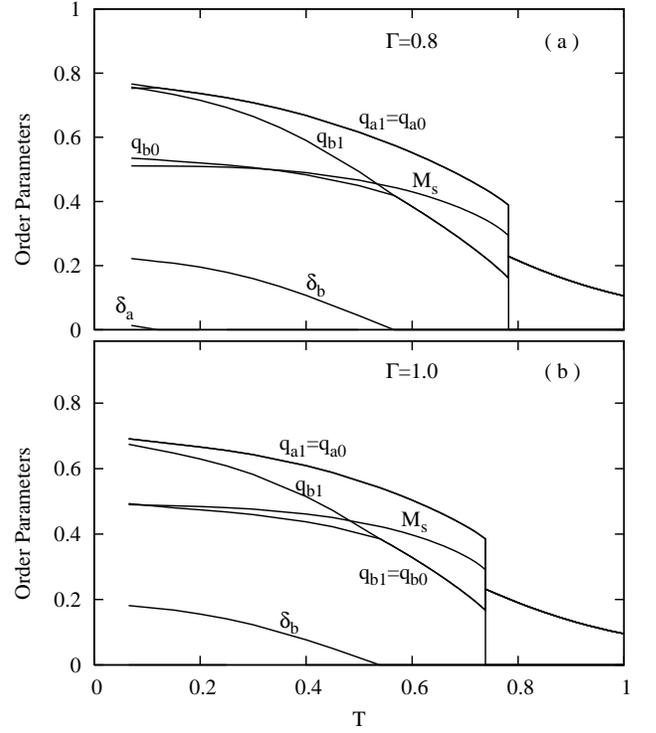} 
 \caption{Order parameters versus $T$ for $H=0.4$,
 $J_0=V_0=0.95$, $J=0$ and $V=1.0$ with: (a)  $\Gamma=0.8$ and (b) $\Gamma=1.0$.}
 \label{paramj0ga8}
\end{figure}

The free energies given by Eqs. (\ref{freeenergy}) and (\ref{freeenergy0}) are compared in the insert of Fig. (\ref{paramj0ga0}). 
The difference between them is very small if it is compared with the difference obtained by the RS and the RSB solution with $m_a=m_b$ (see Fig. (\ref{params4free})).
In this sense, we could use the approximation $m=m_a=m_b$ for the 1S-RSB solution with good results within the asymmetric RSB region for the case with $J\neq0$.

Considering $\Gamma=0.0$, the $H$-dependence of the 1S-RSB order parameters is exhibited in Fig. (\ref{paramj0ga0t04}) for two values of temperature: $T=0.35$ and $T=0.3$.
The RSB occurs in both sublattices at $H=0$. 
When $H$ increases, $\delta_b$ enhances and $\delta_a$ decreases in the asymmetric RSB region.
At $T=0.35$ (Fig. (\ref{paramj0ga0t04}-a)), $\delta_a$ becomes zero within an $H$ range where only the sublattice $b$ presents RSB solution. Therefore, the $H$ field can decrease the temperature in which the sublattice $a$ exhibits RSB.
However, the effects of $H$ are not strong enough to produce a stable RS solution for a particular sublattice at lower temperatures, 
as it is shown in Fig. (\ref{paramj0ga0t04}-b) for $T=0.3$ where $\delta_a$ and $\delta_b$ are greater than zero in the mixed phase.

\begin{figure*}[t!]
\includegraphics[angle=270,scale =0.52]{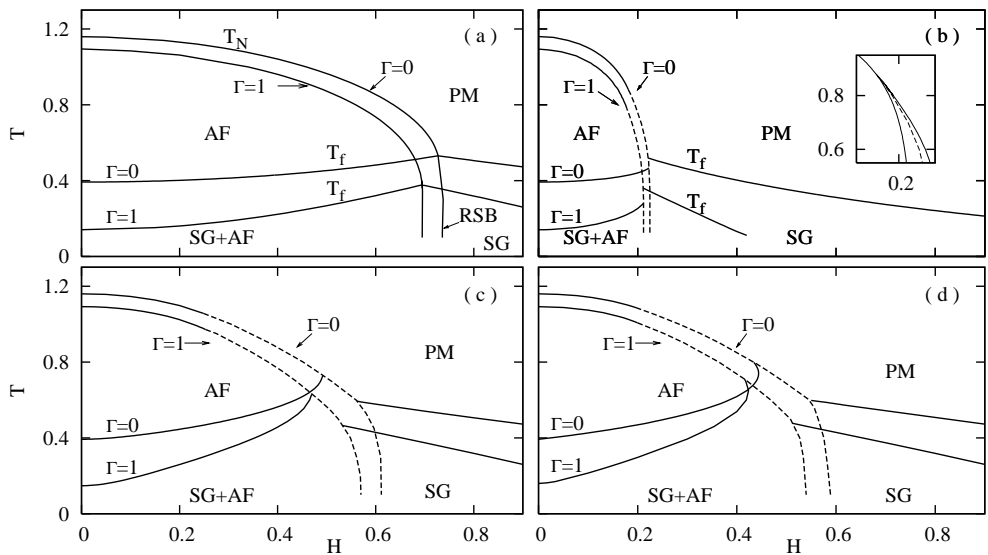} 
\caption{Phase diagrams $T$ versus $H$ when $\Gamma=0$ and $\Gamma=1$ for the following 
interactions: (a) $J=V$ and $V_0=J_0=0.95$, (b) $J=V$, $V_0=1.56$ and $J_0=0.34$, (c) $J=0.24$, $V=0.97$ and $V_0=J_0=0.95$, and (d) $J=0$, $V=1$ and $V_0=J_0=0.95$.
The full (dashed) lines correspond to second-order (first-order) phase transitions.
The insert exhibits the stability limit for the PM and the AF solutions when $\Gamma=0$.}
\label{figveqj}
\end{figure*}

The effects of $\Gamma$ on the order parameters are shown in Fig. (\ref{paramj0ga8}) for $H=0.4$, $\Gamma=0.8$ and $\Gamma=1.0$. 
The RSB region  of the sublattice $a$ decreases faster than that of the sublattice $b$ when $\Gamma$ is applied in the asymmetric RSB region. Hence,
for $\Gamma$ sufficiently high ($\Gamma=1.0$), the RSB solution appears only in sublattice $b$, while sublattice $a$ presents RS solution in all range of $T$ analyzed in Fig. (\ref{paramj0ga8}-b).
Although these results ($\Gamma>0$) are limited by the static approximation, which is expected to yield inaccurate quantitative results at very low temperature, they could provide some qualitative evidences about the RSB behavior at least for $T>0$.\cite{zimmerprb}
For example, when the strength of $\Gamma$ is enhanced, quantum fluctuations assume a relevant role inducing tunneling between the metastable valleys that characterize the SG free energy landscape. 
It could reduce the importance of the RSB, but it is not able to restore the RS solution in the sublattice $b$ for instance.
The quantum fluctuations decrease the region where both sublattices present RSB solution.
However, they can increase 
the region where only sublattice $b$ presents RSB. In this case, the quantum fluctuations can decrease the number of frustrated sites in the asymmetric RSB region.

\subsection{Phase Diagrams}\label{diagrama}
After a detailed study of the asymmetric RSB region (mixed phase), 
we analyze phase diagrams $T$ versus $H$ for the two types of coupling previously discussed.
The main effects of the quantum fluctuations on the phase boundaries and the role of intra-sublattice disordered interactions on phase diagrams are presented.
First, we consider the thermodynamic behavior of the phase boundaries within a region where the RS solution is stable (at $T>T_f$). 
In this case, analytical expressions for the free energy and 
the order parameters are achieved by assuming $q_{p}\equiv q_{p0}=q_{p1}$ with $m=0$ ($\sigma_{1p}=0$ in Eq. \ref{variance}).
The transition from the PM to the AF order can be characterized by the onset of the staggered magnetization $M_s$.
If this transition is continuous, we can expand the set of saddle-point equations in powers of $M_s$. After some lengthy calculations, we find a compact result:
\begin{equation}
M_s=A M_s+B M_{s}^{3}+C M_{s}^5+\cdots
\label{expansao}
\end{equation}
where the coefficients $A$ and $B$ are defined in Appendix \ref{apendicea}.
A second-order phase transition from the PM to the AF order is found when the condition $A=1$, $B<0$ and $C<0$ is obtained.
Tricritical points can be found by the condition $A=1$ and $B=0$ with $C<0$.
We do not calculate the coefficient $C$ and so we will not check the condition $C<0$.

The phase diagrams for interactions as cases (i), (ii), (iii) and (iv) are shown in Figs. (\ref{figveqj}-a), (\ref{figveqj}-b), (\ref{figveqj}-c) and (\ref{figveqj}-d), respectively. 
They exhibit a general feature: the AF order is destroyed by $H$ at the same time that the mixed phase enhances when $H$ increases toward $H_c$. 
Here, $H_c$ is the value of the parallel magnetic field  in which $T_f$ meets $T_N$.
For $H>H_c$, the system behaves as in the one-lattice model, where 
$T_f$ decreases monotonically with $H$.\cite{sk} 
Particularly, for the case (i), the phase boundaries are second-order and the phase diagram (Fig. (\ref{figveqj}-a)) is similar to those one obtained in Ref. \onlinecite{zimmer} for the KS model.

The role of the intra-sublattice parameters
$V_0$ and $V$ on the phase diagrams can be  analyzed by comparing the cases (i) and (ii) (Figs. (\ref{figveqj}-a) and (\ref{figveqj}-b)), and the cases (i), (iii) and (iv) (Figs. (\ref{figveqj}-a), (\ref{figveqj}-c) and (\ref{figveqj}-d)), respectively.
In case (ii), $V_0$ is increased at the same time that $J_0$ decreases in order to keep $\eta$  ($T_N(H=0)/T_f(H=0)$) constant. 
As a consequence, the AF region reduces and a first-order phase transition appears (see Fig (\ref{figveqj}-b)). 
A tricritical point is obtained by expansion (\ref{expansao}). The first-order boundary is located choosing the solution which minimizes the free energy.
The transition from the RS solution to the RSB solution has a discontinuity at $H_c$, where $T_f$ jumps when it crosses from the region $H<H_c$ to $H>H_c$.
In this case, a discontinue transition from the AF order to the SG phase raises when $H$ increases.
Particularly for $\Gamma=0$, the average internal field  $\bar{h}_{p}$ ($=H+V_0 M_p-J_0 M_{p^{'}}$) that acts on the sublattice $p$ is directly affected by the relation between $V_0$ and $J_0$. 
The $\bar{h}_{p}$ in which the PM solution becomes stable is achieved with smaller $H$ when 
$V_0/J_0$ enhances.  
Furthermore, the $\bar{h}_p$ for a stable PM solution can be obtained with an $H$ in which the AF solution is still energetically favorable (see inserts of Fig. (\ref{figveqj}-b)), 
where the first-order transition occurs.
There are numerical evidences within 1S-RSB ansatz which indicate that the mixed/SG phase transition can also be first-order (see Fig. (\ref{params4h}-b)).
Therefore, a first-order transition can occur when $V_0/J_0$ is sufficiently large.

The intra-sublattice disorder also affects the effective AF interaction between the two sublattices as we can see by comparing Fig. (\ref{figveqj}-a) with Figs. (\ref{figveqj}-c) and (\ref{figveqj}-d). 
For instance, the AF region reduces when the $V$ increases at the same time that $J$ decreases to keep $\sqrt{V^2+J^2}$ constant. 
Thus, a small value of $H$ is able to destroy the AF order. 
Besides, 
as shown in previous sections, the degree of RSB depends on the ration $V/J$.
For example, $\delta_b$ increases with $V/J$ in the mixed phase when $H>0$. 
Consequently, the derivative $|\partial{T_f}/\partial{H}|$ is higher in the AF/mixed phase transition when $V$ increases (see Figs. (\ref{figveqj}-c) and (\ref{figveqj}-d)).
Case (iii) shows a phase diagram similar to that one in case (iv). However, the RSB properties of these two cases are very distinct as it has been shown in sections (\ref{resultaA}) and (\ref{resultaB}).
A first-order phase transition also appears for the cases (iii) and (iv).
However, different from Fig. (\ref{figveqj}-b), the $T_f$ in the asymmetric region (AF/mixed phase transition) is higher than in the symmetric region (PM/SG transition) for $H$ close to $H_c$. Consequently, a  discontinue transition from the mixed phase to the PM phase can occur with the increase of $H$ (Figs. (\ref{figveqj}-c) and (\ref{figveqj}-d)).
Therefore, the intra-sublattice ferromagnetic average interaction can introduce a first-order phase transition in this two-sublattice problem. On the other hand, the intra-sublattice SK-type of disorder changes the behavior of $T_f$ that presents a higher increment with $H$ in the AF order when this disorder enhances.

The internal field is even more complex when $\Gamma>0$. 
The critical lines and the tricritical points are decreased when $\Gamma$ increases, as we can see in Fig. (\ref{figveqj}) for phase diagrams with $\Gamma=1.0$. 
The quantum effects are more pronounced at lower temperatures. 
Particularly, the behavior of $T_f$ is changed by the quantum fluctuations. In this case, the derivative $|\partial{T_f}/\partial{H}|$ is increased by $\Gamma$ when $H$ is close to $H_c$. The region of discontinue transition from the AF order to the SG phase (Fig. (\ref{figveqj}-b)) and from the mixed phase to the PM  phase (Figs. (\ref{figveqj}-c) and (\ref{figveqj}-d)) are also increased by $\Gamma$. This $T_f$ behavior occurs because of the competition between quantum fluctuations (dominant at lower $T$) and the asymmetry caused by $H$ (important near the $H_c$) that depends on the intra- and inter-sublattices disordered interactions as previously shown.

\section{Summary and Conclusions}
We studied a fermionic version for the two-sublattice 
Ising SG model in a transverse ($\Gamma$)
and a parallel ($H$) magnetic fields.
This model allows SK-type interactions for spins of same sublattices and between spins of distinct sublattices. The problem is formulated in a Grassmann path integral formalism within the static approximation.
The disorder is treated by the replica trick with Parisi's scheme of one-step replica symmetry breaking (1S-RSB).
A given degree of frustration $\eta=\sqrt{V^2+J^2}/(V_0+J_0)$ is used to analyze the main role of each parameter of the model.
Particularly, we studied the effects of the quantum fluctuations on the asymmetric RSB region and on phase diagrams.

For $H=0$, the two sublattices show symmetric results for the 1S-RSB order parameters, which are identical for all configuration of intra- and inter-sublattices interactions that satisfy the same $\eta$. However, the $H$ field introduces an asymmetric RSB region, in which the  sublattices exhibit different 1S-RSB results that depend on the configuration of intra- and inter-sublattices disorders.
For instance, the degree of RSB of sublattice $b$ ($\delta_b$) is higher than $\delta_a$. 
This difference and the value of $\delta_b$ increase when the intra-sublattice disorder ($V$) enhances.
Nevertheless, for a general case, the RSB occurs at the same temperature for both sublattices when the inter-sublattice interactions are disordered ($J>0$).
In addition, the staggered magnetization has a weak dependence on $T$ in the mixed phase for 1S-RSB solution. 
Results with $\Gamma>0$ suggest that the quantum tunneling decreases the effects of the RSB for both sublattices, but it is not able to restore the RS solution at $T\rightarrow 0$. 
This result is in agreement with that one of fermionic one-lattice quantum SG problem.\cite{zimmerprb} 

For a particular case with intra-sublattice  disordered interactions and only antiferromagnetic coupling 
between the sublattices ($J= 0$), the RSB appears in each sublattices independently. The sublattice $b$ can present RSB solution while the sublattice $a$ still presents RS solution in the asymmetric RSB region. However, the ground state of both sublattices always occurs with RSB solution for $\Gamma=0$. 
Nevertheless, the results within 1S-RSB suggest that the frustration in the mixed phase is changed by the $\Gamma$ field. In this case, the quantum fluctuations affect with more intensity the RSB region of the sublattice $a$ than the sublattice $b$. The sublattice $a$ can exhibit a stable RS solution at any finite temperature with the RSB solution present in the sublattice $b$ for a strong enough $\Gamma$ within the asymmetric RSB region. 

The phase diagrams
show a general feature in this work: a transition from the PM to the AF order then to a RSB region when temperature $T$ decreases for low $H$. The RSB occurs at the freezing temperature $T_f$ with a finite staggered magnetization $M_s$ that characterizes a mixed phase. 
A transition from the PM to the SG phase (RSB solution with $M_s=0$) is observed for high $H$.
The $H$ field destroys the AF phase, but it can favor the RSB region within a certain range.
In this range, $T_f$ increases with $H$. The intra-sublattice interaction affects the $T_f(H)$ behavior that presents a higher increase with $H$ when the intra-sublattice disorder enhances. The presence of an intra-sublattice ferromagnetic long range interaction can introduce a first-order phase transition from the RS to the RSB solution when $H$ increases. 
The $\Gamma$ field always suppresses the magnetic orders. 
However, its effects on $T_f(H)$ are more evident at lower values of $H$, where the asymmetry between the sublattices is small.

\appendix
\section{Expansion for the Neel Temperature}\label{apendicea}
The continuous phase transition from the PM phase ($M_s=0$) 
to the AF order ($M_s>0$) can be obtained by expanding the saddle-point equations 
in powers of $M_s$. 
After some calculations, we find expansion (\ref{expansao}) 
where 
\begin{equation}
A=\beta(V_0+J_0)f^{10}_{0}+\beta^2(V^2-J^2)(q_1f^{01}_{0}+x_1f^{00}_{1})
\label{coefA}
\end{equation}
and
\begin{equation}
B=\beta(V_0+J_0)B_1+\beta(V_0-J_0)B_2 +\beta^2(V^2-J^2)B_3,
\label{coefB}\end{equation}
with
\begin{equation}
\begin{split}
B_1=\beta^2(V_0+J_0)^2\frac{f^{30}_{0}}{6}
 +\beta(V_0+J_0)
 \beta^2(V^2-J^2)
 \\
\times [q_1\frac{f^{21}_{0}}{2} +x_1\frac{f^{20}_1}{2}]
 +\beta^4(V^2-J^2)^2[q_1x_1f^{11}_1
 \\+q_1^2\frac{f^{12}_0}{2} +x_1^2\frac{f^{10}_2}{2}]
 + \beta^2(V^2+J^2)[q_2f^{11}_0+x_2f^{10}_1],
\end{split}
\label{b1}
\end{equation}
\begin{equation}
\begin{split}
B_2=m_2\left\{ \beta(V_0+J_0)f^{20}_0 + \beta^2(V^2-J^2)\right.
 \\
 \left.\times[q_1f^{11}_0+x_1f^{10}_1]\right\},
\end{split}
\label{b2}
\end{equation}
\begin{equation}
\begin{split}
B_3= q_3f^{01}_0 +x_3f^{00}_1  +\beta^2(V^2+J^2)
  [q_1(x_2f^{01}_1 
  \\
  +q_2f^{02}_0) +x_1(q_2f^{01}_1 +x_2f^{00}_2)]
  +\beta^4(V^2-J^2)^2
 \\ \times
 [q_1^3\frac{f^{03}_0}{6}
 +x_1^3\frac{f^{00}_3}{6}
 +q_1x_1^2\frac{f^{01}_2}{2}+x_1q_1^2\frac{f^{02}_1}{2}],
\end{split}\label{b3}
\end{equation}
%
$M\equiv (M_a+M_b)/2=m_2M_s^2+O(M_s^4)$, $q_s\equiv (q_{a}-q_{b})/2=q_1M_s+q_3M_s^3+\cdots$, $q\equiv (q_a+q_b)/2=q_2M_s^2+O(M_s^4)$, $\chi_s\equiv (r_a-r_b+q_b-q_a)/2=x_1M_s+x_3M_s^3+\cdots$ and $\chi\equiv (r_a+r_b+q_a+q_b)/2=x_2M_s^2+O(M_s^4)$. In Eqs. (\ref{coefA}), (\ref{b1}), (\ref{b2}) and (\ref{b3})
\begin{equation}
f^{ij}_{l}=\int Dz \frac{\partial^{(i+2j)} }{\partial y^{(i+2j)}}g_l(z,y)
\end{equation}
where $i,j,l=0,1,2$ or 3, and
\begin{equation}
g_0(z,y)=\frac{\int D\xi\frac{\partial}{\partial y}K}{\Theta},
\end{equation}

\begin{equation}
g_1(z,y)=\frac{\int D\xi\frac{\partial^2}{\partial y^2}K}{\Theta}
-\frac{\int D\xi\frac{\partial}{\partial y}K\int D\xi K}{\Theta^2},
\end{equation}
%
\begin{equation}
\begin{split}
g_2(z,y)=\frac{\int D\xi\frac{\partial^4}{\partial y^4}K}{\Theta}
-2\frac{\int D\xi\frac{\partial^2}{\partial y^2}K \int D\xi\frac{\partial}{\partial y}K}{\Theta^2}
\\
+\int D\xi K[-\frac{\int D\xi\frac{\partial^3}{\partial y^3}K}{\Theta^2}
+2\frac{(\int D\xi\frac{\partial}{\partial y}K)^2}{\Theta^3}],
\end{split}\end{equation}
%
\begin{equation}
\begin{split}
g_3(z,y)=\frac{\int D\xi\frac{\partial^6}{\partial y^6}K}{\Theta}
-3\frac{\int D\xi\frac{\partial^4}{\partial y^4}K \int D\xi\frac{\partial}{\partial y}K}{\Theta^2}
\\
+3\int D\xi\frac{\partial^2}{\partial y^2}K [-\frac{\int D\xi\frac{\partial^3}{\partial y^3}K}{\Theta^2}
+2\frac{(\int D\xi\frac{\partial}{\partial y}K)^2}{\Theta^3}]
\\
+\int D\xi K [-6\frac{(\int D\xi\frac{\partial}{\partial y}K)^3}{\Theta^4} 
-\frac{\int D\xi\frac{\partial^5}{\partial y^5}K}{\Theta^2}
\\
+6\frac{\int D\xi\frac{\partial}{\partial y}K \int D\xi\frac{\partial^3}{\partial y^3}K}{\Theta^3}],
\end{split}\end{equation}
\begin{equation}
 K=\frac{y\sinh(\sqrt{y^2+\beta^2\Gamma^2})}{\sqrt{y^2+\beta^2\Gamma^2}}
\end{equation}
with $y=\beta h$ and $\Theta$ defined in Eq. (\ref{difference}) for $h=h_p$ in the RS solution ($\delta_{1p}=0$) and the PM phase ($M_a=M_b$, $q_a=q_b$ and $r_a=r_b$).

\begin{acknowledgments} This work is supported by the Brazilian agency CNPq.\end{acknowledgments}

\end{document}